\documentclass[twoside]{article}
\usepackage{Proc_NTSE}

\begin{document}
\thispagestyle{plain}

\begin{center}
{\Large \bf \strut
%Insert the title of your contribution here
 Infrared and ultraviolet cutoffs in variational calculations with a harmonic oscillator basis
\strut}\\
\vspace{10mm}
{\large \bf 
%Insert the authors here. Use upper indexes a, b, c, etc., to bind authors with their addresses
% as shown below.
Sidney A Coon$^{a}$}
\end{center}

\noindent{% Insert the addresses  here.
\small $^a$ \it Department of Physics, University of Arizona, Tucson, Arizona USA \\}

%The next command defines running titles:
\markboth{
%Put here the list of authors that will be displayed in running titles:
Sidney A Coon}
{%Put here the short title of your contribution that will be displayed in running titles:
Infrared and Ultraviolet Cutoffs} 

\begin{abstract}

I abstract from a recent publication \cite{Coon2012} the motivations for, analysis in and conclusions of a study of the
 ultraviolet and infrared momentum regulators induced by the necessary truncation of the model spaces formed by  a variational trial wave function.  This trial function is built systematically from  a complete set of many-body basis states  based upon three-dimensional harmonic oscillator (HO) functions.  Each model space is defined by  a truncation of the expansion characterized by a counting number ($\mathcal{N}$) and by  the intrinsic scale  ($\hbar\omega$) of the HO  basis.
   Extending both the uv cutoff to infinity and the ir cutoff to zero  is prescribed for a converged calculation.  In \cite{Coon2012}  we  established practical procedures which utilize these regulators to obtain the extrapolated  result from sequences of calculations with model spaces.   Finally, I update this subject  by mentioning  recent work on our extrapolation prescriptions which have appeared since the submission of \cite{Coon2012}.   The numerical example chosen for this contribution consists of  calculations of the ground state energy of the triton with  the  ``bare" and ``soft" Idaho N$^3$LO nucleon-nucleon ($NN$) interaction.
\\[\baselineskip] 
{\bf Keywords:} {\it no-core shell model; convergence of expansion in harmonic oscillator functions; ultraviolet regulator, infrared regulator}
\end{abstract}

\section{Introduction}
\label{intro}
The advent of giant nuclear shell-model codes based upon the three-dimensional harmonic oscillator (HO) in the 1970s coincided with  the advent of a program to use the HO eigenfunctions as a basis of a finite linear expansion to make a straightforward variational calculation of the properties of light nuclei \cite{Moshinsky}. At the same time theorems based upon functional analysis established the asymptotic convergence rate of these latter  calculations as a function of the counting number (call it $\mathcal{N}$) which characterizes the size of the expansion basis (or model space) \cite{Delves72, Sch72}.  The convergence rates  of these theorems (inverse power laws in $\mathcal{N}$ for ``non smooth" potentials with strong short range correlations and exponential in  $\mathcal{N}$ for ``smooth" potentials such as gaussians) were demonstrated numerically in \cite{Delves72} for the HO expansion and in \cite{Fabre} for the parallel expansion in hyperspherical harmonics.  These convergence theorems seem to be known in the hyperspherical harmonic community and are effectively demonstrated in the calculation of  the properties of few-nucleon systems ``from first principles"; that is, solving the many-body Schr\"{o}dinger equation with a Hamiltonian containing nucleon-nucleon interactions fitted to scattering data and to properties of the deuteron.  The convergence rates of variational calculations using the HO basis have been periodically rediscovered empirically by those who, in the present day, have adapted ``giant shell-model codes" or written new codes to perform ``{\itshape ab initio}" ``no-core shell model" (NCSM) calculations of $s$- and $p$-shell nuclei. I have never seen a reference to the functional analysis theorems  regarding these convergence rates in the NCSM papers.  However, the HO expansion basis has  an intrinsic scale parameter $\hbar\omega$ which does not naturally fit into an extrapolation scheme based upon $\mathcal{N}$ as discussed by \cite{Delves72, Sch72, Kvaal}.  Indeed the model spaces of these NCSM approaches are characterized by the ordered pair ($\mathcal{N},\hbar\omega$).   Here the basis truncation parameter $\mathcal{N}$ and the HO energy parameter $\hbar\omega$ are variational parameters \cite{NavCau04, Maris09,NQSB}.  It is the purpose of this contribution to summarize the properties  of another ordered pair which perhaps more physically describes the nature of the model spaces and provides extrapolation tools which use  $\mathcal{N}$ and $\hbar\omega$ on an equal footing \cite{Coon2012}.  
This is the pair of ultraviolet (uv) and infrared (ir) cutoffs (each a function of both $\mathcal{N}$ and $\hbar\omega$) induced by the truncation.  They were first introduced to the NCSM in \cite{EFTNCSM} in the context of an effective field theory (EFT) approach (for a recent review of this program see  \cite{Ionel_Jimmy}) . These cutoffs or regulators  can usefully be employed in novel extrapolation schemes \cite{Coon2012} which are a natural outgrowth of those  introduced in the 1970s, rediscovered by the NCSM community, and  in current use.

The variational approach alluded to above generates a trial wave function  in a completely systematic manner without regard for the details of the Hamiltonian under consideration other than the implementation of exact symmetries.  The goal, then, is to define a complete set of states for a few-body system and to construct and diagonalize the Hamiltonian matrix in a truncated basis of these states. The result of the diagonalization is an upper bound to the exact eigenvalue of the complete set.  With this method a reliable estimate of the accuracy attained can be made with the variational upper bound \cite{Delves72} provided that the trial function is constructed using the terms of a systematic expansion set and convergence of the diagonalization result (such as a ground-state energy) is observed as the basis is increased.  The algebra appropriate to generating and using trial wave functions, based on three dimensional HO eigenfunctions, has been given by Moshinsky \cite{Moshinsky} and others \cite{others}.  The trial functions take the form of a finite linear expansion in a set of known functions 
 \[\Psi_T = \sum_{\nu}a_\nu^{({\mathcal{N}})}h_\nu  \]
 where $a_\nu^{({\mathcal{N}})}$ are the parameters to be varied and  $h_\nu$ are many-body states based on a summation over products of HO functions.   The advantage of a HO basis is that it is relatively straightforward to construct a complete set of few-body functions of appropriate angular momentum and symmetry; examples are given in \cite{others,JLS70}.
 The trial function must have a definite symmetry reflecting the  composition of the bound state: fermions or bosons.   This trial function $\Psi_T$ must be quadratically integrable and the expectation value of the Hamiltonian must be finite.   The expansion coefficients (known as generalized Fourier coefficients in the mathematical literature) 
depend on the upper limit (such as an $\mathcal{N}$ defined in terms of total oscillator quanta) and are obtained by minimizing the expectation value of the  Hamiltonian in this basis. 
Treating the coefficients $a_\nu^{({\mathcal{N}})}$ as variational parameters in the Rayleigh quotient  \cite{Kruse}, one performs the variation by diagonalizing the many-body Hamiltonian in this basis.
This is an eigenvalue problem so the minimum with respect to the vector of expansion coefficients always exists and one obtains a bound on the lowest eigenvalue (and indeed on the higher eigenvalues representing the excited states \cite{Don33}).  The basis functions can also depend upon a  parameter (such as the harmonic oscillator energy $\hbar\omega$ which sets a scale)  that then becomes a  non-linear variational parameter additional to the linear expansion coefficients.

One can view a shell-model calculation as a variational calculation, and thus expanding the configuration space merely serves to improve the trial wave function \cite{Irvine}.  The traditional shell-model calculation involves trial variational wave functions which are linear combinations of Slater determinants.  Each Slater determinant corresponds to a configuration of $A$
fermions distributed over $A$ single-particle states.  If we take any complete set of orthonormal single-particle wave functions and consider all possible $A$-particle Slater determinants that can be formed from them, then these  wave functions form a complete orthonormal set of wave functions spanning the $A$-particle Hilbert space.  The Slater determinant basis of HO single-particle wave function is often defined in the ``$m$-scheme".   That is, the single-particle states
are labelled by the quantum numbers $n$, $l$, $j$, and $m_j$, where $n$ and $l$ are the radial and orbital HO quantum numbers, $j$ is the total single-particle spin, and $m_j$ its projection along the $z$-axis.  The many-body basis states have well-defined total spin projection, which is simply the sum of $m_j$ of the single-particle states   $M_j = \sum m_j$, hence the name ``$m$-scheme". The many-body basis states  are limited only by the imposed symmetries --- parity, charge 
and total angular momentum projection ($M$), as well as by $\mathcal{N}$.  However, in general the many-body basis states do not have a well defined total $J$.  This scheme  is simple to implement and in two calculations (for positive and negative parity) one gets the complete low-lying spectrum, including the ground state, even though the spins of the low-lying states are not specified in the trial wave function.  The truncation by   $\mathcal{N}$ results in finite matrices to be diagonalized, but they are much larger than the matrices of the Moshinsky program which  expects the properties of the trial wave function ($JT$ basis in shell model language) to be known.  However, because these  shell model wave functions do span the space, an expansion in such ``$m$-scheme" Slater determinants is, in principle, also capable of giving an exact representation of the eigenfunctions of the Hamiltonian.

 These early {\itshape ab initio} calculations, both of the ``no-core" shell model in which all nucleons are active \cite{Irvine}  and of the Moshinsky program \cite{Ciofi85,Portilho02}  attempted to overcome the challenges posed by ``non-smooth" two-body potentials  by including Jastrow type two-body correlations in the trial wave function.   Nowadays, the $NN$ potentials are tamed by unitary transformations within the model space \cite{OSL} or in free space by the similarity renormalization group evolution \cite{Jurgenson}.  In both cases, this procedure generates effective many-body interations in the new Hamiltonian.  Neglecting these destroys the variational aspect of the calculation (and the physics contained in the calculation, of course).  We  retain the variational nature of our NCSM investigation  by choosing a realistic smooth nucleon-nucleon interaction Idaho N$^3$LO \cite{IdahoN3LO}  which has been used previously without renormalization for  light nuclei($A \leq$ 6) \cite{NavCau04}. This potential  is inspired by chiral perturbation theory and fits the two body data quite well.
 It is  composed of contact terms and irreducible pion-exchange expressions multiplied by a regulator function designed to smoothly cut off high-momentum components in accordance with the low-momentum expansion idea of chiral perturbation theory.   The version we use has the high-momentum cutoff of the regulator set at 500 MeV/$c$.  The Idaho N$^3$LO potential is a rather soft one, with heavily reduced high-momentum components as compared to earlier realistic $NN$ potentials with a strongly repulsive core.  Alternatively, in coordinate space, the Yukawa singularity at the origin is regulated away so that this potential would be considered ``smooth" by Delves and 
Schneider and the convergence in $\mathcal{N}$ would be expected to be exponential \cite{Delves72, Sch72}.  Even without the construction of an effective interaction, convergence with the Idaho N$^3$LO $NN$ potential is exponential in $\mathcal{N}$, as numerous studies have shown \cite{NavCau04,Jurgenson}.

With the HO basis in the nuclear structure problem,  convergence has been discussed, in practice,  with an emphasis on obtaining those parameters which appear linearly in the trial function (i.e. convergence with $\mathcal{N}$).  Sometimes  for each $\mathcal{N}$ the  non-linear parameter $\hbar\omega$ is varied to obtain the minimal energy  \cite{NavCau04,CP87} for a fixed  $\mathcal{N}$ and then the convergence with $\mathcal{N}$ is examined.   Sometimes $\hbar\omega$ is simply fixed at a value which gives the fastest convergence in $\mathcal{N}$ \cite{JLS70}.  Other extrapolation schemes have been proposed and used \cite{Maris09}.  In all of these schemes, in my opinion, the extrapolation to an infinite basis is effected with the main role played by $\mathcal{N}$ and a secondary role played by  $\hbar\omega$.  The scheme proposed in \cite{Coon2012} gives $\mathcal{N}$ and $\hbar\omega$ equal roles by employing uv and ir  cutoffs which which must be taken to infinity and to zero, respectively to achieve a converged result (see Figure 1).

\begin{figure}[htpb]
\centerline{\includegraphics[width=0.8\textwidth]{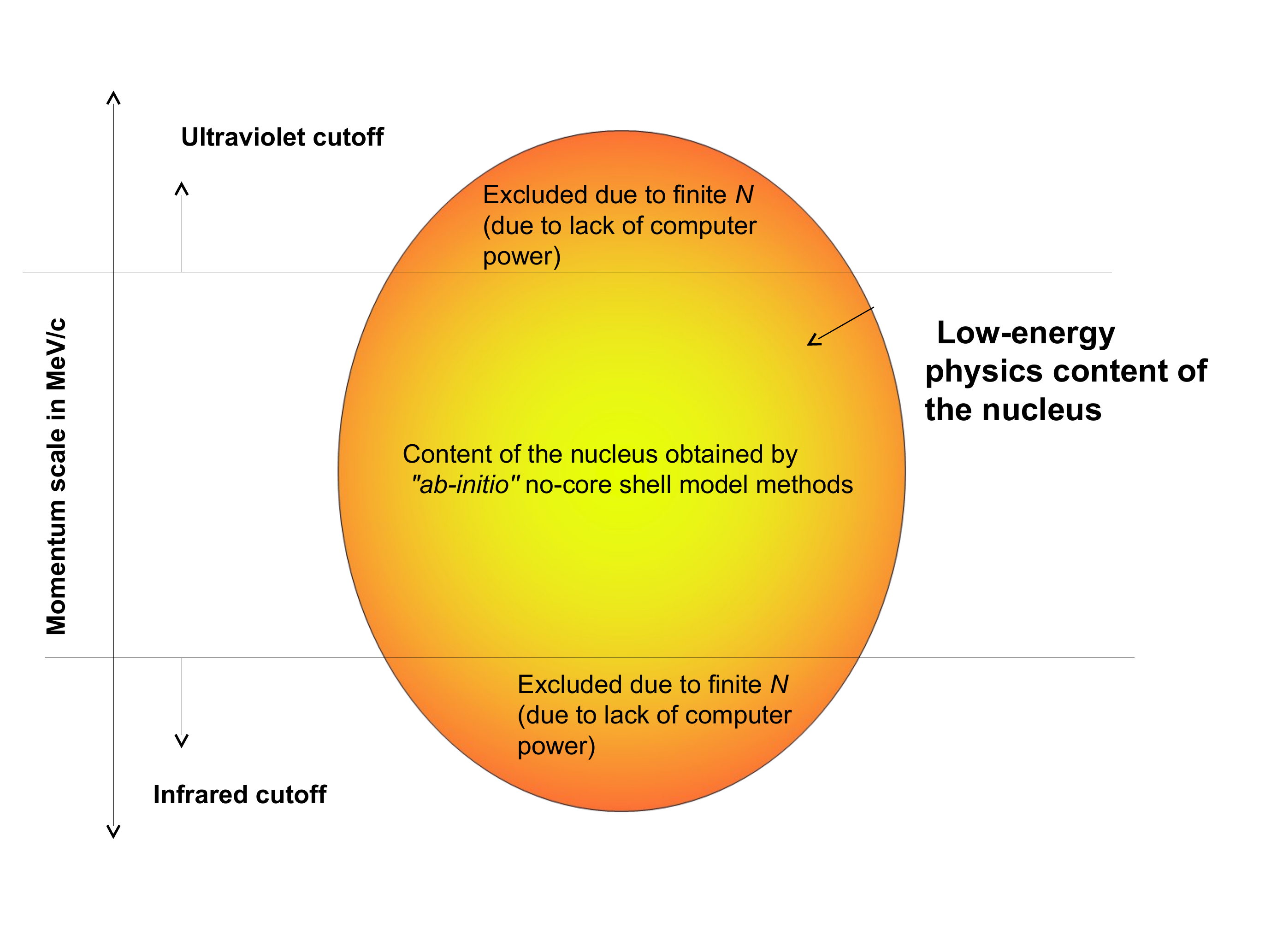}}
\caption{(Color online)  Schematic view of a finite model space (limited by the basis truncation parameter $N$ as described in the text), in which the uv and ir momentum cutoffs are arbitrary.  To reach the full many-body Hilbert space, symbolized by the complete oval,  one needs to let the uv cutoff $\rightarrow \infty$ and the ir cutoff $\rightarrow  0$.}
\label{fig:1}
\end{figure}

In section 2 we briefly describe  expansion schemes in HO functions.   None of the discussion in section 2 is new, but it paves the way for section 3 in which we suggest a convergence analysis based upon the uv and ir cutoffs induced by the truncation of the model space.  Section 4 is devoted to a sampling of tests and examples of this new convergence scheme; for a more extensive discussion with more examples please see Ref.  \cite{Coon2012}.

\section{Expansion in a finite basis of harmonic oscillator functions}
\label{sec1}

 We briefly indicate the workings of the finite HO basis calculations performed and refer the reader to a comprehensive  review article \cite{NQSB} on the no-core shell model (NCSM) for further details and references to the literature. A HO basis allows preservation of translational invariance of the nuclear self-bound system.  Translational invariance is automatic if the radial HO wave function depends on relative, or Jacobi, coordinates as was done in Refs. \cite{JLS70,Ciofi85,Portilho02,CP87}. Antisymmetrization (or symmetrization for the $\alpha$ particle models of \cite{Ciofi85,Portilho02}) of the basis is necessary and described in Refs. \cite{NQSB} and  \cite{NKB}.  Antisymmetrization in a Jacobi basis becomes analytically and computationally forbidding as the number of nucleons increases beyond  four or five.  For this reason  these calculations are alternatively made with antisymmetrized wave functions constructed as Slater determinants of single-nucleon wave functions depending on single-nucleon coordinates. This choice loses translational invariance since, in effect, one has defined a point in space from which all single-particle coordinates are defined. Translational invariance is  restored by choosing a particular truncation  of the basis:  a maximum of the sum of all HO excitations, i.e. $\sum_{i=1}^A (2n_i + l_i) \leq N_{totmax}$, where $n_i,l_i$ are the HO quantum numbers corresponding to the harmonic oscillators associated with the single-nucleon coordinates and $N_{totmax}$ is an example of the generic $\mathcal{N}$ of the Introduction.    The gain of this choice is that one  can use   technology developed and/or adapted for NCSM, such as the shell model code ANTOINE \cite{ANTOINE},  the  parallel-processor codes ``Many-Fermion Dynamics --- nuclear" (MFDn)
\cite{Vary92_MFDn} 
and the No-Core Shell Model Slater Determinant Code \cite{NCSMSD}.  These   codes set up the many-body basis space,
evaluate the many-body Hamiltonian matrix, 
obtain the low-lying eigenvalues and eigenvectors using the
Lanczos algorithm, and evaluate a suite of expectation values using the eigenvectors.  

The eigenstates factorize as products of a wave function depending on relative coordinates and a wave function depending on the c.m. coordinates.  The precise method of achieving the factorization of the c.m. and intrinsic
components of the many-body wave function follows a standard approach, sometimes
referred to as the ``Lawson method" \cite{Lawson}.  
In this method, one selects the many-body basis space in the manner described above 
with $\mathcal{N}=N_{totmax}$
and adds
a Lagrange multiplier term to the many-body Hamiltonian 
$\beta(H_{c.m.} - \frac{3}{2} \hbar\omega)$
where $H_{c.m.}$ is the HO Hamiltonian for the c.m. motion.  With $\beta$
chosen positive (10 is a typical value), one separates the states of lowest c.m. motion 
($0S_{\frac{1}{2}}$)
from the states with excited c.m. motion by a scale of order $\beta \hbar\omega$.  The 
resulting low-lying states have wave functions that then have  the desired factorized form. 
We checked, for the two cases $A=3$ and $A=4$, that the codes $\it manyeff$  \cite{NKB} which use Jacobi coordinates and  No-Core Shell Model Slater Determinant Code  \cite{NCSMSD} based upon single-nucleon coordinates
gave the same eigenvalues for the same values of  $\mathcal{N}=N_{totmax}$ and $\hbar\omega$, indicating that the Lawson method is satisfactory for the calculations in   single-particle coordinates.

Now we return to the truncation parameter $\mathcal{N}$ of the HO basis expansion of the many-body system.  Usually, instead of truncating the sum of all HO excitations $\mathcal{N}=N_{totmax}$, one uses the the more familiar truncation parameter $N_{max}$.
$N_{max}$ is the maximum number of oscillator quanta
shared by all nucleons above the lowest HO configuration for the
chosen nucleus.  
One unit of oscillator quanta is one unit of the quantity $(2n+l)$
where $n$ is the principle quantum number and $l$ is the 
angular quantum number.  For $A=3,4$ systems $N_{max} = N_{totmax}$.   For the $p$-shell nuclei they differ, e.g. for $^6$Li, $N_{max} = N_{totmax}-2$, and for $^{12}$C,  $N_{max} = N_{totmax}-8$.  Later on we will want a truncation parameter which refers, not to the many-body system, but to the properties of the HO single-particle states.
If the highest HO single-particle state of this lowest HO
configuration has $N_0$ HO quanta, then $N_{max}+N_0=N$ identifies the
highest HO single-particle states that can be occupied within this
many-body basis.  Since
$N_{max}$ is the maximum of the {\em total} HO quanta above the
minimal HO configuration, we can have at most one nucleon in such a
highest HO single-particle state with $N$ quanta.  Note that $N_{max}$ characterizes the many-body basis space, whereas $N$ is a label of the corresponding single particle space.  Let us illustrate this distinction with two  examples.  $^6$He is an open shell nucleus with $N_0=1$ since the valence neutron occupies the $0p$ shell in the lowest many-body configuration.  Thus if $N_{max} =4$ the single particle truncation $N$ is 5.  On the other hand, the highest occupied orbital of the closed shell nucleus $^4$He has $N_0=0$  so that $N=N_{max}$.

\section{Ultraviolet and infrared cutoffs induced by basis truncation}

We begin by thinking of the finite single-particle basis space defined by $N$
and $\hbar\omega$ as a model space characterized by two momenta associated with the basis functions themselves.
In the HO basis, we follow \cite{EFTNCSM} and define $\Lambda=\sqrt{m_N(N+3/2)\hbar\omega}$ as the 
momentum (in units of MeV/$c$) associated with the energy of the highest HO level.  The nucleon mass is $m_N=938.92$ MeV. To arrive at this definition one applies the virial theorem to this highest HO level  to establish  kinetic energy as one half the total energy ({\it i.e.}, $(N+3/2)\hbar\omega\:$)  and solves the non-relativistic dispersion relation for $\Lambda$.  This sets one of the two cutoffs for the model space of a calculation.  Energy, momentum and length scales are related, according to Heisenberg's uncertainty principle. The higher the energy or momentum scale we may reach, the lower the length scale we may probe. Thus, the usual definition of an ultraviolet cutoff $\Lambda$ in the continuum has been extended to discrete HO states. It is then quite natural to interpret the behavior of the variational energy of the system with addition of more basis states as the behavior   of this observable with the variation of the ultraviolet cutoff $\Lambda$.  Above a certain value of $\Lambda$ one expects this running of the observable with $\Lambda$ to ``start to behave" so that this behavior can be used to extrapolate to the exact answer.    However, the truncation of the model space by $\mathcal{N}$ implies a second cutoff, absent in free space; an infrared cutoff.  Because the energy levels of a particle in a HO potential are quantized in units of $\hbar\omega$,  the minimum allowed momentum difference between single-particle orbitals  is 
$\lambda=\sqrt{m_N\hbar\omega}$ and that has been taken to be an infrared cutoff \cite{EFTNCSM}.  
That is, there is a low-momentum cutoff $\lambda=\hbar/b$ corresponding to the minimal accessible non-zero momentum (here $b=\sqrt{\frac{\hbar}{m_N\omega}}$ plays the role of a characteristic  length of the HO potential and basis functions).  Note however that there is {\it no} external confining HO potential in place.  Instead the only  $\hbar\omega$ dependence is due to the scale parameter of the underlying HO  basis.  In \cite{EFTNCSM} the influence of the infrared cutoff is removed by extrapolating to the continuum limit, where  $ \hbar\omega\rightarrow 0$    with $N\rightarrow\infty$ so that $\Lambda$ is fixed.  Clearly, one cannot achieve both the ultraviolet limit and the infrared limit by taking $\hbar\omega$ to zero  in a fixed-$N $model space as this procedure takes the ultraviolet cutoff to zero.

 The calculated energies of a many-body system in the truncated model space will differ from those calculated as the basis size increases without limit ($N\rightarrow\infty$).  This is because the system is in effect confined within a finite (coordinate space) volume characterized by the finite value of $b$ intrinsic to the HO basis. The ``walls"  of the volume confining the interacting system spread apart  and the volume increases to the infinite limit  as $\lambda\rightarrow 0$ and $b\rightarrow\infty$ with $\Lambda$ held fixed. Thus it is as necessary to extrapolate the low momentum  results obtained with a truncated basis with a given $b$ or $\hbar\omega$ as it is to ensure that the ultraviolet cutoff is high  enough for a converged result.  These energy level shifts in a large enclosure have long been studied \cite{FukudaNewton}; most recently with the explicit EFT calculation of a triton in a cubic box allowing the edge lengths to become large (and the associated  ir cutoff due to momentum quantization in the box going towards zero) \cite{Kreuzer}.  There it was shown that as long as the infrared cutoff was small compared to the ultraviolet momentum cutoff appearing in the ``pionless" EFT, the ultraviolet behavior of the triton amplitudes was unaffected by the finite volume.  More importantly, from our point of view of desiring extrapolation guidance, this result means that calculations in a finite volume can confidently be applied to the infinite volume (or complete model space) limit. Similar conclusions can be drawn from the ongoing studies of systems of two and three nucleons trapped in a HO potential with interactions from pionless EFT  combined with this definition of the infrared cutoff ($\lambda=\sqrt{m_N\hbar\omega}$); see the review \cite{Ionel_Jimmy}.

Other studies define the ir cutoff as  the infrared momentum which corresponds to the maximal
radial extent  needed to encompass the many-body system we are attempting to describe by the finite basis space (or model space).  These studies find it natural to define the ir cutoff by $\lambda_{sc}=\sqrt{(m_N\hbar\omega)/(N+3/2)}$ \cite {Jurgenson,Papenbrock}.  Note that $\lambda_{sc}$ is the inverse of the root-mean-square (rms) radius of the highest single-particle state in the basis; $\langle r^2\rangle^{1/2}=b
\sqrt{N+3/2} $.   We distinguish the two definitions by  denoting the first (historically) definition by $\lambda$ and the second definition by $\lambda_{sc}$ because of its  scaling properties demonstrated in the next Section.

 The extension in \cite{EFTNCSM} of the continuum ultraviolet cutoff to the discrete (and truncated) HO basis with the definition  $\Lambda=\sqrt{m_N(N+3/2)\hbar\omega}$ seems unexceptional.  But, as always when one confidently makes such a statement, there are exceptions.  For example, an effective momentum for a HO state can be defined by
 the asymptotic relation for large $n$ between the radial part $R_{nl}(r)/r$ of the harmonic oscillator functions and the spherical Bessel functions $j_l(kr)$ of radial part of the 3D  plane wave \cite{Bateman}.  Kallio showed that this relation is very accurate at small $r$ for all $n$ values \cite{Kallio}.  The alternate definition, suggested by Vary \cite{Roorkee}, identifies a uv regulator with the ``Kallio momentum" defined by this relation so that $\Lambda_{alternate} = \sqrt{2}\Lambda$.  This is a scale change only as is the definition by fiat in \cite{FHP} which arrives at the same $\sqrt{2}$ factor for  their $\Lambda$.   The more important distinction is the alternate 
 definitions of the ir cutoff which have different functional forms.   It is clear that increasing $\Lambda$   by increasing $\hbar\omega$ in a fixed-$N$ model space is not sufficient; doing so   increases both of the putative infrared cutoffs as well because  $\Lambda = \lambda\sqrt{N+3/2} = \lambda_{sc}(N+3/2)$ and one continues to effectively  calculate in an effective confining volume which is getting smaller rather than larger. This confining volume is certainly removed by letting  $N\rightarrow\infty$, at fixed $\hbar\omega$, because HO functions form a basis of the complete space. In addition, taking $N\rightarrow\infty$ simultaneously removes the uv cutoff defined by $\Lambda$ and the ir cutoff defined either by  $\lambda$ or $\lambda_{sc}$.   But increasing $N$ without limit  is computationally  prohibitive.  Thus there is a practical issue to address:  whether one must take the ir cutoff to zero by taking $\hbar\omega\rightarrow 0$  at fixed $\Lambda$  ($\lambda_{ir}\equiv\lambda$ definition) or whether it is sufficient to allow $\hbar\omega$ be some larger value, perhaps near that used in  traditional shell-model calculations, and let an increasing $N$ take $\lambda_{ir}$ to small values, as it does with the definition $\lambda_{ir}\equiv \lambda_{sc}$.
 
\section{A study of uv and ir cutoffs in the triton}

We display in a series of figures the running of the ground-state eigenvalue of a single nucleus, $^3$H, on the truncated HO basis by holding one cutoff of ($\Lambda,\lambda_{ir}$) fixed and letting the other vary.   These $^3$H calculations were made for $N\leq 36$ and  values of $\hbar\omega$ as appropriate for the chosen cutoff value.  For $N\geq 16$, we used the code $\it manyeff$  \cite{NKB}  which uses  Jacobi coordinates and the No-Core Shell Model Slater Determinant Code \cite{NCSMSD}  which use single-particle coordinates for smaller $N$.   We checked that the codes gave the same eigenvalues for overlapping values of $N$, indicating that the Lawson method  satisfactorily restores translational invariance to ground-state energy calculations in   single-particle coordinates. 

\begin{figure}[htpb]
\centerline{\includegraphics[width=0.8\textwidth]{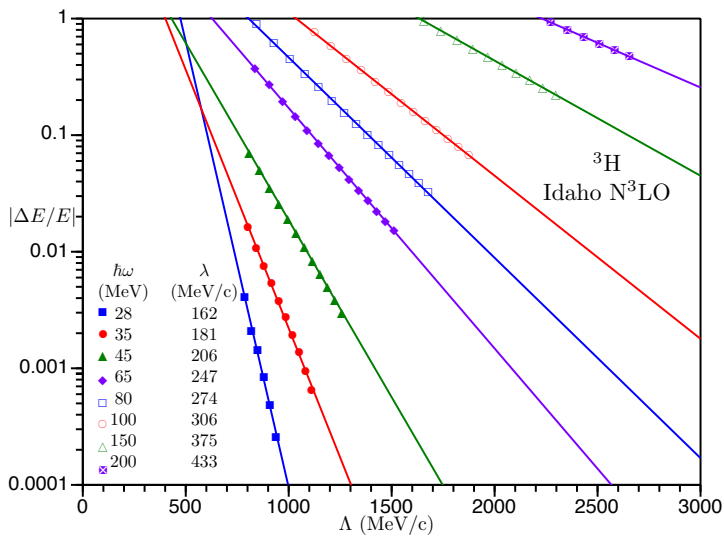}}

\caption{(Color online)  Dependence of the ground-state energy of $^3$H (compared to a converged value; see text) upon the uv momentum cutoff   $\Lambda=\sqrt{m_N(N+3/2)\hbar\omega}$ for different fixed $\lambda=\sqrt{m_N\hbar\omega}$.  The curves are fit to the calculated points.}

\label{fig:2}
\end{figure}

In Figure 2 and the following figures,  $\vert\Delta E/E\vert $ is defined as $\vert (E(\Lambda,\lambda_{ir}) - E)/E\vert$ where $E$ reflects a consensus ground-state energy from benchmark calculations with this $NN$ potential,  this nucleus, and different few-body methods. The accepted value for the ground state of  $^3$H with this potential is $ -7.855$ MeV from a 34 channel Faddeev calculation \cite{IdahoN3LO}, $ -7.854$ MeV from a hyperspherical harmonics expansion \cite{Kievsky08}, and $-7.85(1)$ from a NCSM calculation \cite{NavCau04}.

For the choice of Figure 2, $\lambda_{ir}\equiv \lambda=\sqrt{m_N \hbar\omega}$, 
$\vert\Delta E/E\vert $ decreases exponentially at fixed $\lambda$, as $\Lambda$ increases for the values of $\Lambda$ achieved in this study.   Fixed $\hbar\omega$ implies $N$   $\it{alone}$ increases to drive $\Lambda\rightarrow\infty$,  $\lambda_{sc}\rightarrow 0$ simultaneously.   The linear fit on a semi-log plot is extracted from the data. 
 For fixed $\Lambda$, a smaller $\lambda$ implies a smaller 
 $\vert\Delta E/E\vert$ since more of the infrared region is included in the calculation.
 
%\pagebreak
\begin{figure}[htpb!]
\centerline{\includegraphics[width=0.8\textwidth]{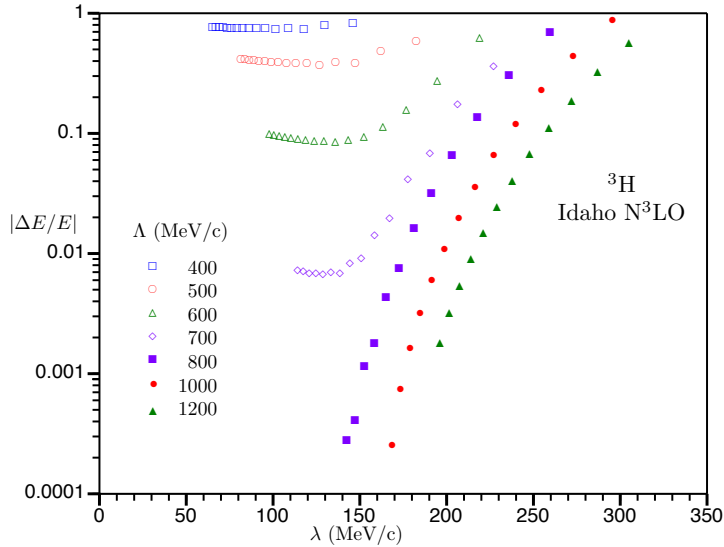}}
\caption{(Color online)  Dependence of the ground-state energy of $^3$H (compared to a converged value; see text) upon the ir momentum cutoff $\lambda=\sqrt{m_N\hbar\omega}$ for fixed  $\Lambda=\sqrt{m_N(N+3/2)\hbar\omega}$.  }
\label{fig:3}
\end{figure}

In Figure 3 we hold fixed the uv cutoff of ($\Lambda,\lambda_{ir}$) to display the running of  $\vert\Delta E/E\vert$ upon the suggested ir cutoff $\lambda$.   For fixed $\lambda$, a larger $\Lambda$ implies a smaller  $\vert\Delta E/E\vert$ since more of the uv region is included in the calculation.  
But we immediately see a qualitative change in the curves between the transition $\Lambda=700$ MeV and $\Lambda=800$ MeV; for smaller $\Lambda$, $\vert\Delta E/E\vert$ does not go to zero as the ir cutoff is lowered and more of the infrared region is included in the calculation.  This behavior suggests that $\vert\Delta E/E\vert$ does not go to zero unless $\Lambda\geq\Lambda^{NN}$, where
 $\Lambda^{NN}$ is some uv regulator scale of the $NN$ interaction itself.  From this figure one estimates  $\Lambda^{NN}\sim$ 800 MeV/$c$ for the Idaho N$^3$LO interaction. 
 
 Yet the description of this interaction in the literature says that the version we use has the high-momentum cutoff of the regulator set at $\Lambda_{N3LO} = 500$ MeV/$c$ \cite{IdahoN3LO}.  This does not mean that the interaction has a sharp cutoff at exactly  500 MeV/$c$, since the terms in the Idaho N$^3$LO interaction are actually regulated by an exponentially suppressed term of the form 
 
 \[  \exp  \left [ - \left( \frac{p}{\Lambda_{N3LO}}\right) ^{2n} - \left( \frac{p'}{\Lambda_{N3LO}}\right) ^{2n}\right] .    \]
In this expression, $p$ and $p'$ denote the magnitude of the  initial and final nucleon momenta of this non-local potential in the center-of-mass frame and $n\geq 2$.  Because the cutoff is not sharp, it should not be surprising that one has not exhausted the uv physics of this interaction for values of single-particle $\Lambda$ somewhat greater than 500  MeV/$c$.  Note that this form of the regulator allows momentum transfers ($\vec p - \vec p'$) to achieve values in the range up to $2  \Lambda_{N3LO}$.   Can one make an  estimate of the uv regulator scale of the Idaho N$^3$LO interaction which is more appropriate to the discrete HO basis of this study?  An emulation of this interaction in a harmonic oscillator basis uses  $\hbar\omega = 30$ MeV and $N_{max} = N=20$  \cite{Barnea2}. Nucleon-nucleon interactions are defined in the relative coordinates of the two-body system so one should calculate $\Lambda^{NN} =\sqrt{m (N + 3/2)\hbar\omega} $ with the $\it reduced$ mass $m$ rather than the nucleon mass $m_N$ appropriate for the single-particle states of the model space.  Taking this factor into account, the successful emulation of the  Idaho N$^3$LO interaction in a HO basis suggests that $\Lambda^{NN}\sim$ 780 MeV/$c$, consistent with the figure.

For $\Lambda < \Lambda^{NN}$ there will be missing contributions of size $\vert (\Lambda-\Lambda^{NN})/\Lambda^{NN}\vert$  so ``plateaus" develop as $\lambda\rightarrow 0$,  revealing this missing contribution to $\vert\Delta E/E\vert$. We cannot rule out the possibility of a plateau appearing at the level of 0.0001 or less for $\Lambda\geq 800$ MeV/$c$ as $\lambda\rightarrow 0$.  This is because the smallest $\lambda$ available to our calculations is limited by 
$\lambda = \Lambda/\sqrt{N+3/2}$ and the largest $N=36$ with our computer resources.  That is, the
leftmost calculated points of Figure 3 move to higher values of $\lambda$ as fixed $\Lambda$ increases above 800 MeV/$c$.  At fractional differences of 0.001 or less, the development of possible plateaus could be masked by round-off errors in the subtraction of two nearby numbers, each of which may have its own error.  Nevertheless, the ``plateaus" that we do see  are not flat as $\lambda\rightarrow 0$ and, indeed, rise significantly with decreasing $\Lambda <\Lambda^{NN}$.  This suggests that corrections are needed to  $\Lambda$ and $\lambda$ which are presently defined only to leading order in $\lambda/\Lambda$.  The authors of \cite{FHP} take our suggested simile of a truncated basis to a confining region  quite seriously and use it to obtain a  first order correction to both $\Lambda$ and $\lambda_{ir}$.  We hope to learn if higher-order  corrections  can be directly determined by our data in a future study.

\begin{figure}[htpb]
\centerline{\includegraphics[width=0.8\textwidth]{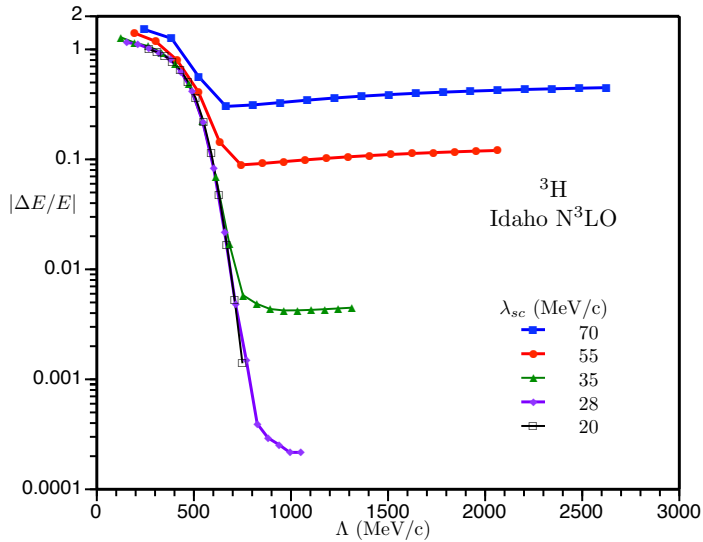}}
\caption{(Color online)  Dependence of the ground-state energy of $^3$H (compared to a converged value; see text) upon the uv momentum cutoff   $\Lambda=\sqrt{m_N(N+3/2)\hbar\omega}$ for different values of  the ir momentum cutoff $\lambda_{sc}=\sqrt{(m_N\hbar\omega)/(N+3/2)}$. Curves are not fits but simple point-to-point line segments to guide the eye. }
\label{fig:4}
\end{figure}

Now we turn to the second pair of cutoffs  of ($\Lambda,\lambda_{ir}$) and display in Figure 4 the analogue of Figure 2 except that this time  $\lambda_{ir}\equiv \lambda_{sc}  =  \sqrt{m_N \hbar\omega/(N + 3/2)} $.   
For fixed $\lambda_{sc}$, $\vert\Delta E/E\vert$ does not go to zero with increasing $\Lambda$, and indeed even appears to rise for  fixed $\lambda_{sc}\geq 35$ MeV/$c$ and $\Lambda\geq 800$ MeV/$c$. Such a plateau-like  behavior was attributed in Figure 3  to a uv regulator scale characteristic of the $NN$ interaction.  Can the  behavior of Figure 4 also be explained by  a  ``missing contributions" argument; i.e. an argument based upon $\lambda_{sc}\leq\lambda^{NN}_{sc}$ where  $ \lambda^{NN}_{sc}$ is a second characteristic ir regulator scale implicit in the $NN$ interaction itself?  One can envisage such an ir cutoff as related to the lowest energy configuration that the $NN$ potential could be expected to describe.  For example, the inverse of the $np$ triplet scattering length of 5.42 fm corresponds to a low-energy cutoff of about 36 MeV/$c$.   The previously mentioned  emulation of the Idaho N$^3$LO interaction in a harmonic oscillator basis \cite{Barnea2}  has $\lambda^{NN}_{sc}\sim$ 36 MeV/$c$.  At low $\Lambda$ and $\lambda_{sc}\leq \lambda^{NN}_{sc}$, $\vert\Delta E/E\vert$ does fall with increasing $\Lambda$ and this behavior can be fitted by a Gaussian as shown for $^3$H and  and other $s$-shell nuclei  in \cite{Coon2012}.   But we will see in the next figure that one has not yet captured the uv region at these low values of $\Lambda$.

\begin{figure}[htpb]
\centerline{\includegraphics[width=0.8\textwidth]{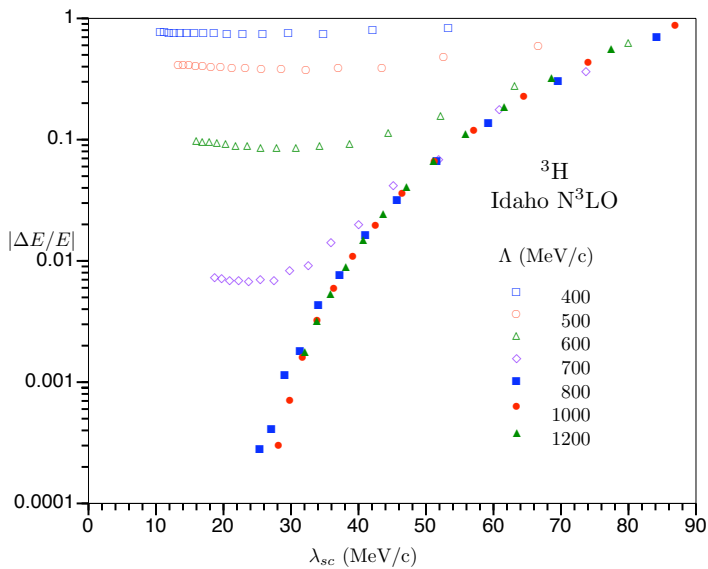}}
\caption{(Color online)  Dependence of the ground-state energy of $^3$H (compared to a converged value; see text) upon the ir momentum cutoff $\lambda_{sc}=\sqrt{(m_N\hbar\omega)/(N+3/2)}$ for fixed  $\Lambda=\sqrt{m_N(N+3/2)\hbar\omega}$. }
\label{fig:5}
\end{figure}

Figure 5 is the analogue to Figure 3: only the variable on the x-axis changes from $\lambda$ to $\lambda_{sc}  =  \lambda^2/\Lambda$.  For $\Lambda < \Lambda^{NN}\sim 780$ MeV/$c$ the missing contributions  and resulting ``plateaus" are as evident as in Figure 3. (Please see discussion of Figure 3 for an account of possible ``plateaus" for larger values of $\Lambda$.) The tendency of these plateaus to rise as $\lambda_{sc}\rightarrow 0$ again suggests
 a refinement is needed to this first-order definition of the cutoffs.   Around  $\Lambda\sim 600$ MeV/$c$ and above the plot of $\vert\Delta E/E\vert$ versus $\lambda_{sc}$ in Figure 5  begins to suggest a universal pattern, especially at large $\lambda_{sc}$.
For $\Lambda\sim 800$ MeV/$c$ and above the pattern defines a universal curve for all values of  $\lambda_{sc}$.  This is  the region  where $\Lambda\geq\Lambda^{NN}$indicating that nearly all of the ultraviolet physics set by the potential has been captured.
Such a universal curve suggests that $\lambda_{sc}$ could be used for extrapolation to the ir limit, provided that $\Lambda$ is kept large enough to capture the uv region of the calculation.  Figure 5 is also the motivation for our appellation $\lambda_{sc}$, which we read as ``lambda scaling", since  this figure exhibits the attractive scaling properties of this regulator.

\begin{figure}[htpb]
\centerline{\includegraphics[width=0.8\textwidth]{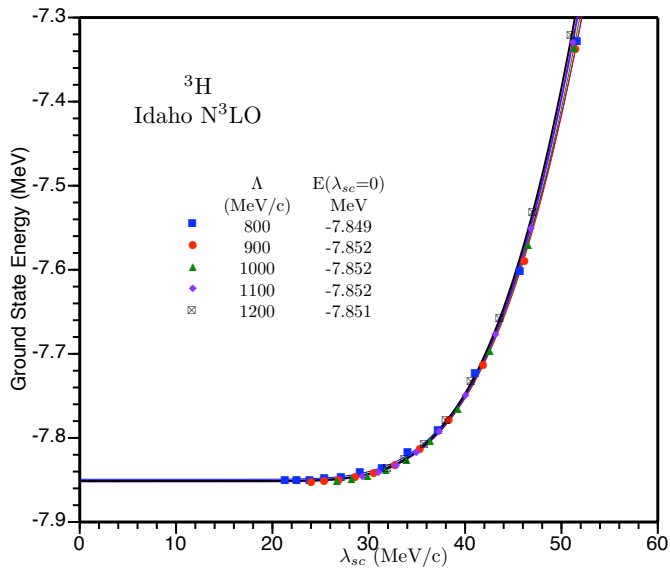}}
\caption{(Color online) The ground state energy of $^3$H calculated at five fixed values of   $\Lambda=\sqrt{m_N(N+3/2)\hbar\omega}$ and variable $\lambda_{sc}=\sqrt{(m_N\hbar\omega)/(N+3/2)}$. The curves are fits to the points and the functions fitted are used to extrapolate to the ir limit $\lambda_{sc}= 0$. }
\label{fig:6}
\end{figure}

We now utilize the scaling behavior displayed on Figure 5  to suggest an extrapolation procedure which we demonstrate in Figure 6. The extrapolation is performed by a fit of an exponential plus a constant to each set of results at fixed  $\Lambda$.  That is, we fit the ground state energy with three adjustable parameters using the relation   \( E_{gs}(\lambda_{sc}) = a \exp(-b/\lambda_{sc}) + E_{gs}(\lambda_{sc}=0)  \). The mean and standard deviation of the five values of $E_{gs}(\lambda_{sc}=0)  $ were 
 $-7.8511$ MeV  and  0.0011 MeV, respectively, as suggested by Figure 7  in which the overlap of the five separate curves cannot be discerned. 
 It should be noted that our five extrapolations in Figure 7 employ an exponential function whose argument
$1/\lambda_{sc}  =  \sqrt{(N + 3/2)/(m_N \hbar\omega)}$ is proportional to $\sqrt{N/(\hbar\omega)}$.
This extrapolation procedure of taking $\lambda_{sc}  =  \sqrt{m_N \hbar\omega/(N + 3/2)}$ toward the smallest value allowed by computational limitations treats both $N$ and $\hbar\omega$ on an equal basis.  The exponential extrapolation in $\sqrt{N/(\hbar\omega)}$ is therefore distinct from the popular extrapolation which employes an exponential in $N_{max}$ ($=N$ for this $s$-shell case) \cite {NavCau04,  Maris09, NQSB, Jurgenson} and provides a refinement to the procedures of the 1970s for dealing with ``smooth" potentials.

This extrapolation procedure treats both $N$ and $\hbar\omega$ on an equal basis.  For example, the extrapolation at fixed $\Lambda  =1200$ MeV/$c$ employs values of $\hbar\omega$ from 41 to 65 MeV and $N=22-36$.  The one at fixed $\Lambda  =800$ MeV/$c$ employs values of $\hbar\omega$ from 18 to 44 MeV  and   $N=14-36$. The curves of Figure 6  encompass values of $\lambda_{sc}$ between 20 and 52 MeV/$c$.   We attempted to
quantify the spread in extrapolated values by  fitting only segments of the curves of this figure.  Recall that the smallest value of $\lambda_{sc}$ requires the largest $N$.  Fits to the   segment from $\lambda_{sc} = 20$ MeV/$c$ to $\lambda_{sc} = 40$ MeV/$c$ (always for the five displayed values of fixed $\Lambda$) resulted in a mean of $-7.8523$ MeV and standard deviation of 0.0008 MeV.  Cutting out the left hand parts of the curves and fitting only  from $\lambda_{sc} = 30$ MeV/$c$ to $\lambda_{sc} = 55$  MeV/$c$ gave a mean of $-7.8498$ MeV and standard deviation of 0.0022 MeV.  For both these trials a rather large $N$ was needed, ranging from 14 to 36 but the extrapolation is quite stable.  In contrast, values of $\lambda_{sc}$ higher than those shown in  Figure 7, namely from $\lambda_{sc} = 50$ MeV/$c$ to $\lambda_{sc} = 85$ MeV/$c$, require fewer computational resources ($N=8-22$).  The extrapolations have a mean  and standard deviation of $-7.792$ MeV and 0.042 MeV, still not so far away from the accepted value of $-7.85$ MeV.

\begin{figure}[htpb]
\centerline{\includegraphics[width=0.8\textwidth]{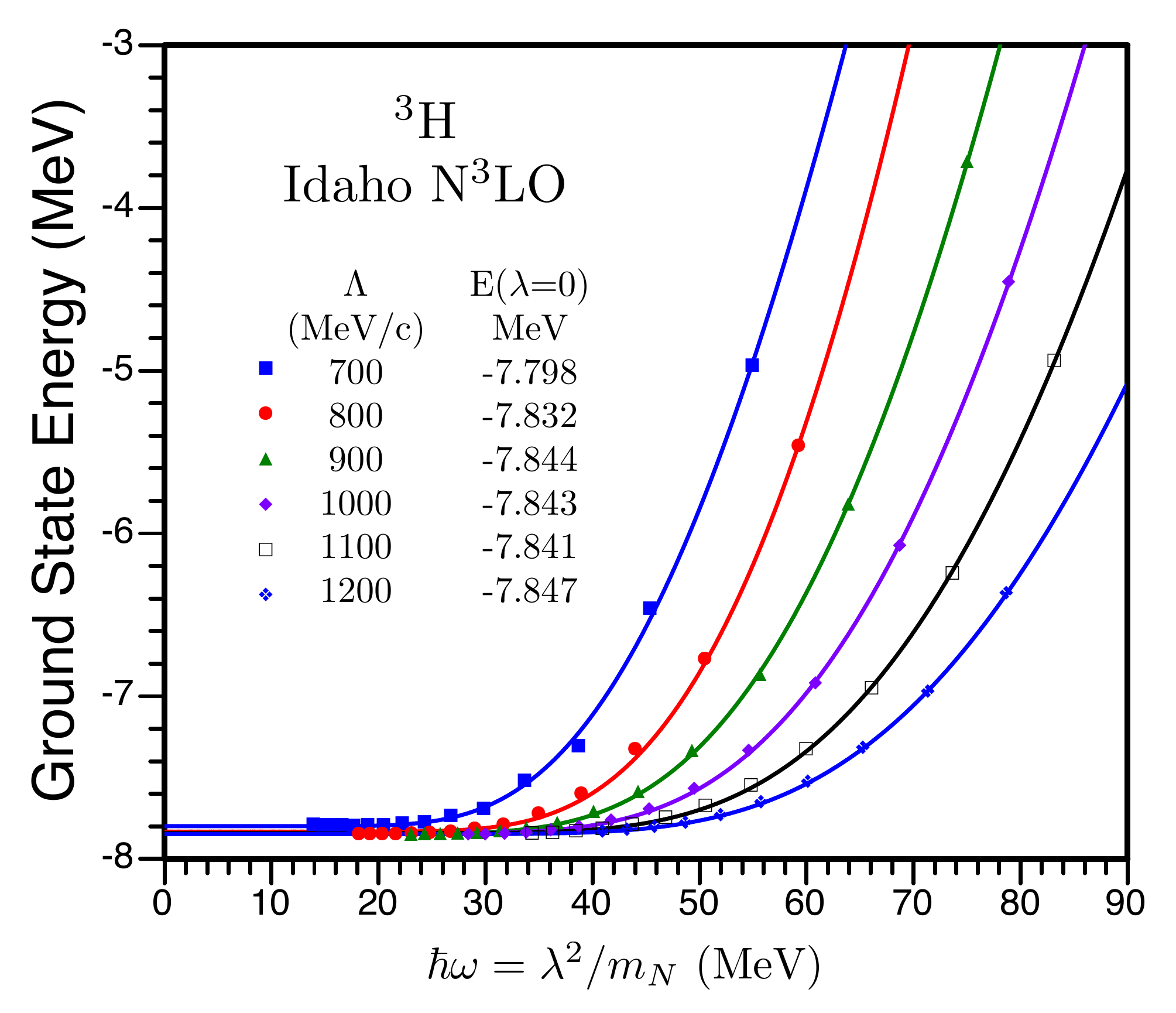}}
\caption{(Color online) The ground state energy of $^3$H calculated at six fixed values of   $\Lambda=\sqrt{m_N(N+3/2)\hbar\omega}$. The curves are fits to the points and the functions fitted are used to extrapolate to the ir limit $\lambda=\sqrt{m_N\hbar\omega} =  0$ with fixed $\Lambda$ as in Figure 6.} 
\label{fig:7}
\end{figure}

Figure 3 suggests that an extrapolation to the infrared limit could equally well  be made by taking $\lambda \rightarrow 0$  for a fixed large $\Lambda$.  Instead we choose to extrapolate in $\hbar\omega$ with an eye to future exploitation of archival calculations made in the variables ($N_{max}$,$\hbar\omega$).   In Figure 7 we fit the ground state energy of $^3$H with three adjustable parameters using the relation 
   \( E_{gs}(\hbar\omega) = a \exp(-c/\hbar\omega) + E_{gs}(\hbar\omega=0)  \) six times, once for each fixed value of $\Lambda$.  It is readily seen that one can indeed make an ir extrapolation  by sending $\hbar\omega\rightarrow 0$ with fixed $\Lambda$ as first advocated in Ref.  \cite{EFTNCSM} and that the five ir extrapolations with $\Lambda > \Lambda^{NN}\sim 780$ MeV/$c$   are consistent.  The spread in the six extrapolated values is about 0.049 MeV or about 1$\%$  about the mean of  $-7.832$ MeV.  The standard deviation is 0.020 MeV.

%\newpage

\begin{figure}[htpb]
\centerline{\includegraphics[width=0.8\textwidth]{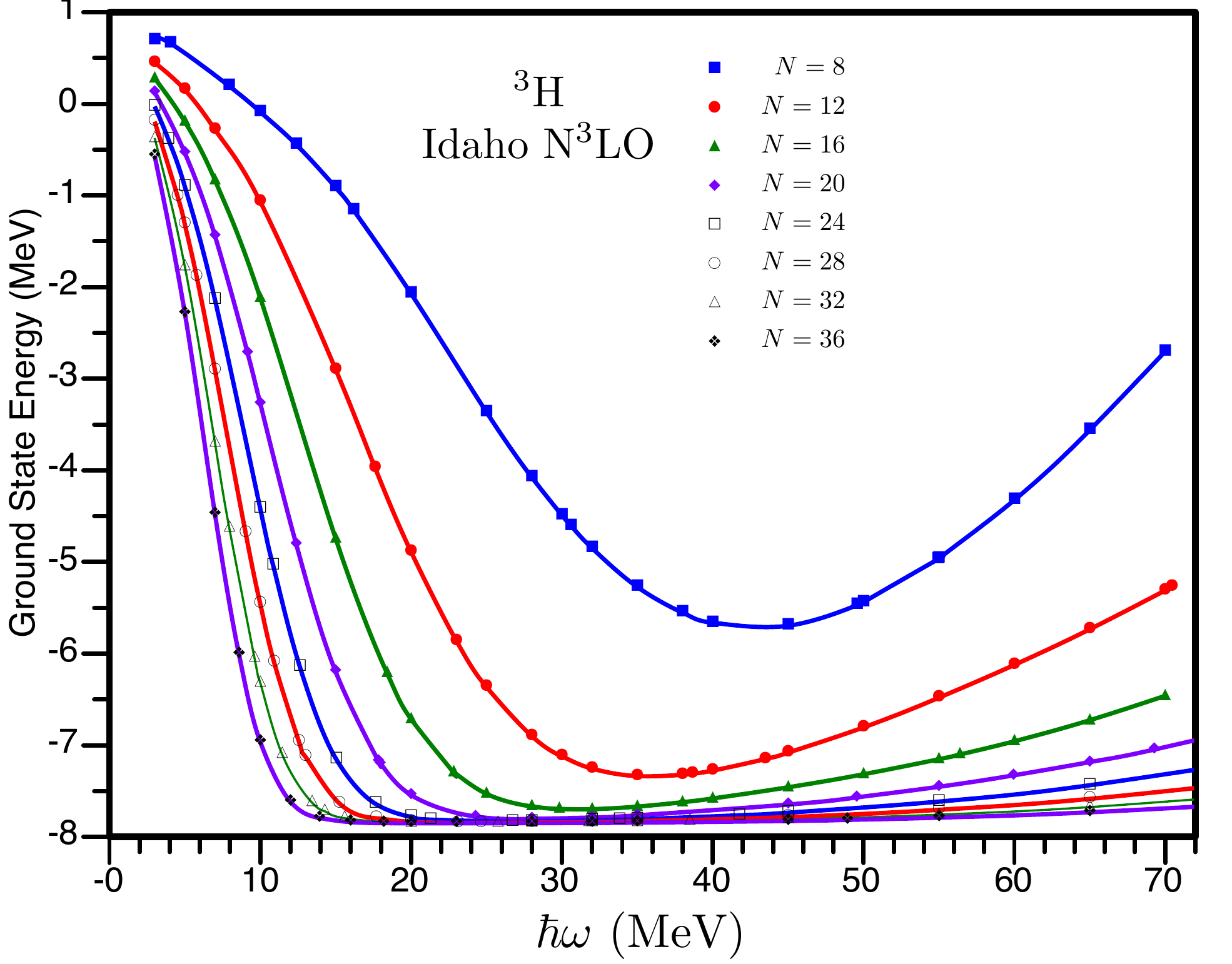}}
\caption{(Color online) Dependence of the ground-state energy of $^3$H upon $\hbar\omega = \lambda^2/m_N = \lambda_{sc}^2/[m_N(N+3/2)]$ for fixed $N = \Lambda^2/\lambda^2 -3/2 = \Lambda/\lambda_{sc} -3/2 $.  Curves are not fits but spline interpolations to guide the eye.  }
\label{fig:8}
\end{figure}

Now let us accept the role of the ordered pair ($\Lambda,\lambda_{ir}$) of cutoffs in these variational calculations and examine the ordered pair ($\mathcal{N},\hbar\omega$).  That is, we take  the basis truncation parameter $\mathcal{N}$ and the HO energy parameter $\hbar\omega$ to be variational parameters.  We now observe convergence as the truncation of the model space is   lessened by increasing $N=N_{max}$, where $N$ is the specific  truncation parameter $\mathcal{N}$ and $N_{max}$ is the total number of energy quanta kept in the basis.  Figure 8 shows a plot of the variational energy of the ground state of $^3$H plotted in this traditional way,  pioneered in Figure 1 of \cite{Moshinsky} and continued through \cite{UNAM} to the present day \cite{Maris09,FHP}.  Optimum values for the parameters that enter linearly can be obtained by solving a matrix eigenvalue problem. But the optimum value of the nonlinear parameter must in principle be obtained by, for example, numerical minimization which could be difficult as the algorithm could easily miss the global minimum and get trapped in a local minima.
The plots such as Figure 8 and others  in the nuclear physics  literature show that 1) for small bases a change in the non-linear parameter $\hbar\omega$ can have a dramatic change in the variational estimate of the ground state energy and 2) the dependence on the nonlinear parameter decreases as the basis size increases.  These observations seem to vitiate the need for an extensive numerical minimization by varying $\hbar\omega $ \cite{caveat}.  For example, in Figure 8 the minimum of each fixed $N$ curve is easily read off the plot.

From Figure 8, we see that the variational energy decreases and thus moves away from the converged value -7.85 MeV as $\hbar\omega \rightarrow 0$ at fixed $N$ (for all $N$ considered!). This is readily understood  in terms of Figure 1.  At fixed  $N$ one captures more  infrared physics by lowering the infrared cutoff ($\lambda_{ir} \propto \sqrt{\hbar\omega}$) but misses the ultraviolet physics because lowering $\hbar\omega$ also lowers the ultraviolet cutoff ($\Lambda \propto \sqrt{\hbar\omega}$).  The loss of uv physics due to the  lower $\hbar\omega$ overwhelms the gain of ir physics and the estimate of the ground state becomes very bad.  A similar situation holds as $\hbar\omega$ increases:  the uv cutoff increases toward $\infty$ so that more uv physics is captured but the ir cutoff also rises and more and more of the infrared physics is lost to the calculation.

The approximate minimum of the $N=8$ curve is at $\hbar\omega\sim 43$ MeV which corresponds to $\Lambda \sim 620$ MeV/$c$ and $\lambda_{sc} \sim 65$ MeV/$c$.   From Figure 5 we  realize that for this small value of  $\Lambda < \Lambda^{NN}\sim 780$ MeV/$c$ and large value of $\lambda_{sc}>\lambda^{NN}_{sc}\sim$ 36 MeV/$c$, we would expect  about a 30\% shortfall in the ground state energy and this is what we see in Figure 8.  At the minimum of the $N=8$ curve the variational parameters are nowhere near their limits in the ($\Lambda,\lambda_{ir}$) regulator picture and the variational energy is not very good.   Because $N \propto \Lambda^2/\lambda^2$ or $N \propto \Lambda/\lambda_{sc}$, increasing the truncation parameter $N$ simultaneously increases the uv cutoff and decreases the ir cutoff so that the curves move lower  and lower. We observe that, as fixed $N$ increases, the minima of each curve moves to a lower value of $\hbar\omega$, as was previously observed in similar calculations for $^4$He with this potential \cite{NavCau04} and for a variety of nuclei ($A=2-16$) \cite{Maris09} with another  realistic $NN$ potential JISP16 \cite{Shirokov07}.  Maris also observes a monotonic movement to the left with a basis truncation on the single-particle basis so that the truncation parameter $\mathcal{N}$ becomes $N_{shell}$ rather than $N_{max}$ \cite{Maris12}.  Apparently another behavior, first a shift to the right and then to the left as fixed $N$ is increased, is noted in \cite{FHP} and interpreted as first an approach to uv convergence and then, as the uv physics is obtained a further convergence in the ir regulator.  We, and other NCSM calculations (including one with a $N_{shell}$ truncation),  do not see this behavior.

In Figure 8, the monotonic movement to a lower 
$\hbar\omega$  is clear as $N$ increases from 8 to 20, all values corresponding to $\Lambda < \Lambda^{NN}\sim 780$ MeV/$c$, the region in which the uv physics has not yet been captured.
As $N$ is increased to $N=24$ (not yet possible  for $p$-shell nuclei with present day computers and codes) the minimum moves down to $\hbar\omega\sim 24$ MeV which corresponds to $\Lambda \sim 790$ MeV/$c$ , $\lambda \sim 150$ MeV/$c$ and $\lambda_{sc} \sim 31$ MeV/$c$.  At these values the uv cutoff seems high enough (see Figures 3 and 5)  and the ir cutoff low enough (see Figures 2 and 4) that one could argue that convergence was nearly reached.   As $N$ increases from 24 to 36 the fixed $N$ curves pile up on each other, but an expanded scale (not shown) separates them to demonstrate that  the minimum stays near 24 MeV ($\Lambda \sim 920$ MeV/$c$ and $\lambda_{sc} \sim 24$ MeV/$c$) and the curves become somewhat independent of $\hbar\omega$ within a limited range.  Even so,  any calculation in a finite basis should be examined from the point of view of the more physical regulators ($\Lambda,\lambda_{ir}$).  This calculation should, in principle, always be extrapolated to the uv and ir limits.
 Independence of $\hbar\omega$ for fixed $N$ is due to a playoff between the uv and ir cutoffs and it should be understood how this playoff  affects the calculation. The often heard mantra ``look for independence upon the value of $\hbar\omega$ because that means the series of calculations has converged" should be retired, in my opinion.

After submission of \cite{Coon2012}, Furnstahl, Hagen and Papenbrock   posted an  investigation of  uv and ir cutoffs in finite oscillator spaces \cite{FHP} .  They assume that $\lambda_{sc}$ (scaled by a factor of $\sqrt{2}$ from the  $\lambda_{sc}$ of this paper) {\it is} the ir cutoff.  They take our suggested simile of a truncated basis to a confining region  quite seriously and use the simile to 
 derive an explicit extrapolation formula in their ir cutoff.   The derived formula  is the same (exponential in $\sqrt{N/(\hbar\omega)}$) as the one of \cite{Coon2012} reviewed here and is used in the same way: establish that the uv cutoff is large enough and then extrapolate in the ir variable.
In addition, they suggest  a first (higher) order correction to both the uv and ir regulators.   The caveat to what they call a ``theoretically derived ir formula"  is the remark made recently by Lieb {\it et al}:  ``If one fixes the particle number N in a very large box and calculates the shift in energy caused by [a given local one-body potential] V, the answer depends on the box shape and boundary conditions" \cite{Lieb}.  But this has always been true \cite{FukudaNewton}. 

As \cite{FHP} assumes that (scaled) $\lambda_{sc}$ is the ir regulator, they took the behavior shown in Figure 4 for small $\lambda_{sc}$ to suggest a second extrapolation formula for the uv cutoff.  That is, 
at low $\Lambda$ and $\lambda_{sc}\leq \lambda^{NN}_{sc}$, $\vert\Delta E/E\vert$  falls with increasing $\Lambda$ and this behavior can be fitted by a Gaussian, as shown for $^3$H and  other $s$-shell nuclei  in \cite{Coon2012}.  This Gaussian  in $\Lambda \propto \sqrt{N(\hbar\omega)}$ then becomes an exponential in $N(\hbar\omega)$.  Their final formula assumes relative independence of the uv and ir extrapolations so it is a sum of  exponentials with arguments proportional to $N(\hbar\omega)$ from the uv regulator and to  $\sqrt{N/(\hbar\omega)}$ from their ir regulator.  These results are a useful advance on the exponential form of convergence in $N$ (with no mention of the role of the scale parameter ${\hbar\omega}$) shown less concretely by  the forty-year-old  theorems of \cite{Delves72} and \cite{Sch72}.  The authors of  \cite{FHP} caution, as have we, that  results such as these should be expected only for the ``smooth" potentials of \cite{Delves72} and \cite{Sch72} (or in their momentum space characterization: ``super-Gaussian falloff in momentum space") such as those inspired by chiral EFT or obtained by renormalization group transformations.   The extrapolation formulae appear to be successful in calculations of  open shell medium to heavy nuclei ($A=74$)  with nuclear interactions inspired by chiral EFT \cite{Soma}.  

There has been a recent turn to consider other bases for expanding the trial wave function;  bases which have a presumed better behavior at large distances than the HO basis which has a Gaussian falloff \cite{desanctis}.  The most effective basis used in few-nucleon physics \cite{Papp},  in nucleon-nucleus scattering \cite{Amos} and in nuclear reactions \cite{Nunes} are the Coulomb-Sturmians.  This is a complete and discrete set of the   eigenfunctions of a Sturm-Liouville problem associated with the Coulomb potential \cite{Weinberg}.  Caprio {\it et al.}  have recently used this basis to make NCSM calculations of light nuclei \cite{Caprio}.  They found it beneficial to link the length scale parameter $b_l$ of the Sturmian with the length scale $b$ of the HO eigenfunction so as to
provide a closer alignment of the low-n Coulomb-Sturmian basis functions with the harmonic-oscillator basis functions.  They choose to {\it formally} truncate the Coulomb-Sturmian basis with an $N_{max}$ counting number.  Thus they end up with the same ordered pair ($\mathcal{N},\hbar\omega$) as with the HO basis.  However,  the $\hbar\omega$  value quoted for the Coulomb-Sturmian basis is simply the  $\hbar\omega$ of the reference oscillator length, from which the actual $l$-dependent length parameters $b_l$ are chosen to align  the low-$n$ Coulomb-Sturmian basis functions with the harmonic-oscillator basis functions. It therefore has no direct significance as an energy scale for the problem.  Moreover, the $N_{max}$ truncation is difficult to interpret as an ``energy cut" as it is for the HO basis.  Caprio {\it et al.} extrapolate to an infinite basis in the following way:  the  non-linear parameter $\hbar\omega$ is varied to obtain the minimal energy   for the highest  $\mathcal{N}$ available,  $\hbar\omega$ is then fixed at that value and  the convergence with $\mathcal{N}$ is assumed to be exponential (extrapolation B of \cite{Maris09}).
This is basically the procedure of Delves \cite {Delves72}, rooted in theorems of functional analysis , and  is not directly related to the EFT inspired cutoffs discussed here.  Given that neither $\mathcal{N}$ nor 
 $\hbar\omega$  are given an energy interpretation in this paper, it is problematic that one can simply take over the arguments of \cite{Coon2012} or \cite{FHP} to define new dimensionful uv or ir cutoffs for use in extrapolation.  Yet the savings in computation and increase in physical understanding should motivate such an effort in the future.

In summary, we have introduced a practical extrapolation procedure with $\Lambda\rightarrow\infty$ and $\lambda_{ir}\rightarrow 0$ which can be used when the size of the HO basis needed exceeds the capacity of the computer resources as it does for $^4$He and $^6$He and certainly will for any more massive nuclei. Unlike other extrapolation procedures the ones advocated in this paper treat the variational parameters $\mathcal{N}$ and $\hbar\omega$ on an equal footing to extract the information available from sequences of calculations with model spaces described  by ($\mathcal{N},\hbar\omega$).   We have established that  $\Lambda$
does not need to be extrapolated to $\infty$ but if   $\Lambda>\Lambda^{NN}$ set by the potential one can make the second  extrapolation to zero with either ir cutoff  $\lambda_{sc}$ (see Figure 6) or $\lambda$ (see Figure 7).  The choice of the scaling cutoff $\lambda_{sc}$ is especially attractive as $\Lambda$ need not be held constant but $\it any$ $\Lambda$ large enough can be used in the ir extrapolation.  The traditional plots in the variables ($\mathcal{N},\hbar\omega$) can be understood by considering the uv and ir cutoffs as primary.

\section{Acknowledgements}
The study culminating in \cite{Coon2012} was conceived  and initiated at the National Institute for Nuclear Theory's program 
{\it Effective field theories and the many-body problem} in the spring of 2009.  This contribution was written while I was enjoying the stimulating hospitality of the fall 2012 INT program {\it Light nuclei from first principles} (INT-PUB-12-052).   I am grateful to my collaborators Michael Kruse and Matthew Avetian for the numerical aspects of the study and include them as well as U. van Kolck and James Vary  for much effort in the interpretive aspects.  I thank Sigurd Kohler for emphasizing to me the importance of the early studies of many-body systems confined to a finite coordinate space volume.

 We are grateful to  Petr Navr\'atil for generously allowing us to use his No-Core Shell Model Slater Determinant Code and his $\it manyeff$ code for  the calculations cited.
The calculations cited  were done with
 allocations of computer time from the UA Research Computing High Performance Computing (HPC) and High Throughput Computing (HTC) at the University of Arizona and from the LLNL
institutional Computing Grand Challenge program.  Numerical calculations
have been performed in part at the LLNL LC facilities supported
by LLNL under Contract No. DE-AC52-07NA27344. This contribution was supported in part by USDOE Division of Nuclear Physics
grant DE-FG02-04ER41338 (Effective Theories of the Strong Interaction) and NSF award 0854912 (New Directions in Nuclear Structure Theory).

%\newpage

%\end{document}

\pagebreak

%\pagebreak

%\pagebreak

\pagebreak

\end{document}